\documentstyle[psfig]{mn}

\newif\ifAMStwofonts
\ifoldfss
  \ifCUPmtlplainloaded \else
    \NewTextAlphabet{textbfit} {cmbxti10} {}
    \NewTextAlphabet{textbfss} {cmssbx10} {}
    \NewMathAlphabet{mathbfit} {cmbxti10} {} % for math mode
    \NewMathAlphabet{mathbfss} {cmssbx10} {} %  "   "    "
  \fi
  \ifAMStwofonts
    \ifCUPmtlplainloaded \else
      \NewSymbolFont{upmath} {eurm10}
      \NewSymbolFont{AMSa} {msam10}
      \NewMathSymbol{\upi}     {0}{upmath}{19}
      \NewMathSymbol{\umu}     {0}{upmath}{16}
      \NewMathSymbol{\upartial}{0}{upmath}{40}
      \NewMathSymbol{\leqslant}{3}{AMSa}{36}
      \NewMathSymbol{\geqslant}{3}{AMSa}{3E}

    \fi
  \fi
\fi % End of OFSS

\ifnfssone
  \newmathalphabet{\mathit}
  \addtoversion{normal}{\mathit}{cmr}{m}{it}
  \addtoversion{bold}{\mathit}{cmr}{bx}{it}
  \newmathalphabet{\mathbfit} % math mode version of \textbfit{..}
  \addtoversion{normal}{\mathbfit}{cmr}{bx}{it}
  \addtoversion{bold}{\mathbfit}{cmr}{bx}{it}
  \newmathalphabet{\mathbfss} % math mode version of \textbfss{..}
  \addtoversion{normal}{\mathbfss}{cmss}{bx}{n}
  \addtoversion{bold}{\mathbfss}{cmss}{bx}{n}
  \ifAMStwofonts
    \ifCUPmtlplainloaded \else
      \UseAMStwoboldmath
      \makeatletter
      \new@mathgroup\upmath@group
      \define@mathgroup\mv@normal\upmath@group{eur}{m}{n}
      \define@mathgroup\mv@bold\upmath@group{eur}{b}{n}
      \edef\UPM{\hexnumber\upmath@group}
      \new@mathgroup\amsa@group
      \define@mathgroup\mv@normal\amsa@group{msa}{m}{n}
      \define@mathgroup\mv@bold\amsa@group{msa}{m}{n}
      \edef\AMSa{\hexnumber\amsa@group}
      \makeatother
      \mathchardef\upi="0\UPM19
      \mathchardef\umu="0\UPM16
      \mathchardef\upartial="0\UPM40
      \mathchardef\leqslant="3\AMSa36
      \mathchardef\geqslant="3\AMSa3E
    \fi
  \fi
\fi % End of NFSS release 1

\ifnfsstwo
  \DeclareMathAlphabet{\mathbfit}{OT1}{cmr}{bx}{it}
  \SetMathAlphabet\mathbfit{bold}{OT1}{cmr}{bx}{it}
  \DeclareMathAlphabet{\mathbfss}{OT1}{cmss}{bx}{n}
  \SetMathAlphabet\mathbfss{bold}{OT1}{cmss}{bx}{n}
  \ifAMStwofonts
    \ifCUPmtlplainloaded \else
      \DeclareSymbolFont{UPM}{U}{eur}{m}{n}
      \SetSymbolFont{UPM}{bold}{U}{eur}{b}{n}
      \DeclareSymbolFont{AMSa}{U}{msa}{m}{n}
      \DeclareMathSymbol{\upi}{0}{UPM}{"19}
      \DeclareMathSymbol{\umu}{0}{UPM}{"16}
      \DeclareMathSymbol{\upartial}{0}{UPM}{"40}
      \DeclareMathSymbol{\leqslant}{3}{AMSa}{"36}
      \DeclareMathSymbol{\geqslant}{3}{AMSa}{"3E}
    \fi
  \fi
\fi % End of NFSS release 2

\ifCUPmtlplainloaded \else
  \ifAMStwofonts \else % If no AMS fonts
    \def\upi{\pi}
    \def\umu{\mu}
    \def\upartial{\partial}
  \fi
\fi

\title{Unbiased reconstruction of the mass function using microlensing survey data.}
\author[C. Alard]
       {C. Alard \\
        Institut d'Astrophysique de Paris, 98bis Boulevard Arago, F-75014.}
\date{}

\pagerange{\pageref{firstpage}--\pageref{lastpage}}
\pubyear{2000}

\begin{document}

\maketitle

\label{firstpage}

\begin{abstract}
The large number of microlensing events discovered towards the Galactic
 Bulge bears the promise to reconstruct the stellar mass function. The
 more interesting issue concerning the mass function is certainly
 to probe its low mass end, near the region occupied by the brown dwarfs.
 However due to the source confusion, even if the distribution
 and the kinematics of the lenses are known, the estimation of the mass 
 function is extremely biased at low masses. The blending due to the source 
 confusion biases the duration of the event, which in turn dramatically
 biases the estimation of the mass of the lens. To overcome this
 problem we propose to use differential photometry of the microlensing events 
 obtained using the image subtraction method. Differential photometry
 is free of any bias due to blending, however the drawback of differential
 photometry is that the baseline flux is unknown. In this paper we will
 show that even without knowing the baseline flux, purely differential
 photometry allow to estimate the mass function without any biases. The
 basis of the method is that taking the scalar product of the microlensing 
 light  curves with a given function and taking its sum over all the 
 microlensing events is equivalent to project the mass function on another 
 function. This method demonstrates that there is a direct correspondancy 
 between the space of the observations and the space of the mass function.
 Concerning the function to use in order to project the observations, we
 show that the principal components of the light curves are an optimal set.
 We also demonstrate that there is no additional information about the 
 distribution of the scalar products of the data beyond their sum (first
 order moment). Higher order moments are only linear combination of the first
 order moment. Thus the sum of the projection 
 on the principal components contains all the information, and translate in
 an equal number of projection of the mass function with functions associated
 with the principal components. To illustrate the method we simulate data
 sets consistent with the microlensing experiments. By using
 1000 of these simulations, we show that for instance
 the exponent of the mass function
 can be reconstructed without any biases. 
\end{abstract}

\section{Introduction.}
 The success of the microlensing experiments has been impressive, several
 collaborations have reported the detection of microlensing amplification
 of stars, OGLE (Udalski {\it et al.} 2000), MACHO (Alcock {\it et al.} 1998),
  EROS (Derue {\it et al.} 1999)), DUO (Alard \& Guibert 1997), 
  VATT (Uglesich  {\it et al.} 1999). In particular, the recent 
 release 
 by the OGLE II collaboration of more than 200 microlensing events is of great
 interest. One of the obvious promises of such data sets is the possibility
 to explore the mass function near the low mass end. Although, the relation
 between the mass function and the microlensing observations is not 
 straightforward. In the classical scheme of analysis, one must first 
 estimate the duration of the events by fitting the theoretical light curves,
 then compute the lensing rates, and finally relate these lensing rates to the
 mass function. The trouble is that fitting the theoretical light curve to the
 microlensing data to obtain the duration of the event is a highly degenerated
 process (Alard 1997, Han 1997, Wozniak \& Paczy\`nski 1997). Due to the source confusion in 
 crowded fields, it is almost impossible to estimate the flux of the amplified
 source reliably. The flux of the source is severely biased by blending with
 neighboring sources. In such case, an over-estimation of the baseline flux 
 will result in an under-estimation of the duration of the event. This bias
 is severe for unresolved stars and will result in a reduction in the
 estimation of the duration of a factor of 2,3, or even more. Since the bias
 on the mass of the lens goes like the square of the bias on the duration, 
 the relevant bias on the mass function will be very large. It was already
 shown (Alard 1997, Han 1997) that the contribution of these unresolved, highly
 blended stars to the lensing rated was dominant at short durations. The 
 consequence is that the estimation of the mass function at low mass end will 
 be largely over-estimated, suggesting the existence of a large number of
 brown dwarf that actually do not exists. It is obvious that as long as
 the bias due to the blended sources has not been solved, any correct
 estimation of the mass function is impossible. This paper proposes a solution
 to this problem. We will see that using differential photometry obtained 
 using the image subtraction method (Alard \& Lupton 1998, Alard 2000)
 it is possible to estimate the mass function, even
 without knowing the baseline flux of the source. This method assumes
 that the distribution and kinematics of the lenses are known with
 good accuracy. It is important to emphasize that the uncertainties related
 to the structure of the Galaxy in the bulge region are several order
 of magnitude smaller than the uncertainties due to the blending bias.
 Many good model of the kinematic and structure of the bulge of our Galaxy
 are already available (Zhao, 1996, Fux 1997, Bissantz {\it et al.}, 1997, Binney, Gerhard, \& Spergel 1997) 

\section{The Method.}
\subsection{Introduction}
 When analyzing microlensing observations one has to deal with the light
 curves of a number N of microlensing candidates. We will assume that we have 
 purely differential photometry only, with a baseline flux equal to zero.
 We emphasize that high quality differential photometry can always
 be obtained after the events have been detected by classical
 methods by using the image subtraction method (Alard 1999).
 These light curves will be represented by the symbol ${\rm \bf S_i(t)}$,
 \ ($j=1,..,N$). 
 Assuming that un-amplified flux of the source is {\bf A},
 the
 expression of the ${\rm \bf S_i(t)}$ as a function of time will be given by
 the theoretical microlensing amplification formula:
\begin{equation}
 S_j(t) = (A-1) \ \frac{u(t)^2+2}{u(t) \ \sqrt{u(t)^2+4}}
\end{equation}
 With:
$$
 u(t) = \sqrt{u_0^2+\left(\frac{t}{t_E}\right)^2}
$$
(we recall that we are working in the assumption that we have at our disposal
 a differential flux only, with for convenience a baseline flux equal to 0). \\\\
 The most important issue is of course how do we relate these observations to the 
 microlensing parameters: 
 the impact parameter, ${\rm \bf u_0}$,
 and the time to cross the Einstein ring: ${\rm \bf t_E}$. And furthermore how do we relate
 the microlensing parameters to the mass function itself. The first serious difficulty
 we encounter is that of course ${\rm \bf u_0, t_E}$, cannot be extracted easily by fitting
 the light curve due to the extreme blending degeneracy 
 (Alard 1997, Han 1997, Wozniak \& Paczynski 1997). Thus our first and most 
essential step
 will be to derive a better set of parameters. This set of parameters will have to be 
 non degenerated with respect to blending. The most simple and most natural way to deal
 with such problem is to express the observations as a linear combination of a small
 number of vectors. The best way to make such a decomposition is known as principal
 components analysis, it is equivalent to an eigen value decomposition. Thus all we have
 to do is to look for the firsts eigen functions, and to 
 calculate the scalar product of the components with our vectors (time series) of
 observations. 
\subsection{Normalization}
 Before making the principal components analysis we have to consider that one
 of the
 parameters, the amplitude of the magnified sources does not contain any information about
 the lensing event. Actually the reason of the the degeneracy of the fit is precisely 
 the unknown amplitude of the source. The only relevant and meaningful parameters are 
  ${\rm {\bf u_0} \ \rm and \ {\bf t_E}}$. Thus it is important to derive an estimator
 which is independent of the amplitude. Since the amplitude is a linear 
parameter, it is sufficient to normalize the vector of observations in order 
to get rid of the 
 amplitude. This normalization can be made in the sense of the scalar 
product we are going to use to make our principal components decomposition. In such
 case, the normalized light curve $\overline S_i(t)$ can be expressed as: \\\\
$$
 \overline S_i(t) = \frac{1}{\alpha} \times \frac{u(t)^2+2}{u(t) \ \sqrt{u(t)^2+4}}
$$
And:
$$
 \alpha = \sqrt{ \int \left[ \frac{u(t)^2+2}{u(t) \ \sqrt{u(t)^2+4}} \right]^2 \ dt}
$$
Note that $\overline A(t)$ does not depend any more on the amplitude, but only on  ${\rm {\bf u_0} \ \rm and \ {\bf t_E}}$. \\\\
 The most simple way to apply the previous normalization to the observations $S_j$ is to calculate directly the modulus of $|S_j|$ from the observations. 
 %$$
 % \overline S_j = S_j/|S_j|
 %$$
 %with:
 %$$
 % |S_j| = \sqrt{\Sigma_I {S^i_j}^2 }  
 %$$
Although it is obvious that calculating directly $|S_i|$ from the observations may not be optimal. To
 estimate $|S_i|$ we will use the following approach: first we fit a microlensing model
 to the observations. Even if the fit is very degenerated, one can always find a good 
 solution which fits the data well. We will call this fitted model $\tilde S_i$. 
 Once we get this light curve solution we estimate using the following formula:
$$
 |S_i| = \sqrt{\int  {\tilde{S_i}(t)}^2 \ dt}
$$
  We will use the same approach for all our other
 calculations, in the same way we will estimate the principal components not by taking
 directly the data, but by taking the models we have fitted to the light curves. And 
 finally the scalar products with the principal components will be also estimated by
 cross products with the theoretical models.
\subsection{Principal components}
As we have already explain, we will first fit a microlensing model $\tilde S_i$ to the light 
curves, normalize this series of N vectors to obtain the series of vectors 
${\overline{ \tilde  S_i}}$,
 and then calculate the principal components of these ${\overline{ \tilde  S_i}}$ vectors. Usually it is possible to express almost all the information with 
 a reduced number of principal components with the following linear 
decomposition:
$$
 {\overline{ \tilde  S_i}} = \sum_j a_{ij} \ P_j(t)
$$
Where $a_{ij}$ is the projection of time series number i, 
 $\bf{ \tilde {S_i} \ (\rm {\bf u_0},\rm {\bf t_E}, \rm {\bf t})}$ on the principal component number j:
$$
  a_{ij} \ (\rm { u_0},\rm { t_E})  = \int \overline{ \tilde {S_i}}\ (\rm { u_0},\rm { t_E},\rm { t}) \ P_j\ (\rm { t}) \ \rm dt
$$
Note that $a_{ij}$ is a function of ${\rm {\bf u_0} \ \rm and \ {\bf t_E}}$:
 $a_{ij} =  \beta \ (\rm { u_0},\rm { t_E})$  \\\\
 For instance a typical set of 100 microlensing light curves can be expressed 
as the combination of only 4 components with an accuracy good to 1 \%. 
The other principal components contains very little additional information, but
 mostly noise.
\subsection{Statistical distribution of the projections on the principal
 components.}
We consider that most of the information is contained in a number ${\rm N_C}$ 
 of principal components. Thus almost all the information can be extracted from
 the distribution of the projection of the $N_1$ time series on the ${\rm N_C}$
 principal components.
The first meaningful quantity concerning the distribution of the $a_{ij}$ is 
 the first order moment of the distribution: \\
$$
 \left<a_{j} \right> = \ \sum_i a_{ij}
$$
The second interesting quantity is the second order moment of the distribution: \\
$$
\left<a_{j}^2 \right> = \ \sum_i a_{ij}^2
$$
But it is important to notice, that $a_{ij}^2 =  \left[ \beta \ (\rm { u_0},\rm { t_E}) \right]^2$, can be written as a scalar product of the data
 with a given function. Since we
 can always find a function $F(\rm t)$ such that:
\begin{equation}
 a_{ij}^2  = \left[ \beta \ (\rm { u_0},\rm { t_E}) \right]^2 = \int \overline{ \tilde { S_i}}\ (\rm { u_0},\rm { t_E},\rm { t}) \ F(t) \ dt
\end{equation}
{\bf Proof: \\}
To prove that $F(t)$ exists we have to show that Eq. (2) can be satisfayed
 for any point in the space of the parameters $(\rm { u_0},\rm { t_E})$. 
 We can map this parameter space by using a regular grid
 with a step as small as we like. This grid will have an almost
 infinite number of rows ($N_u$) and columns ($N_t$). The total number
 of points in the grid is: $N_G=N_u \times N_t$. We define
 $F(t)$ by sampling this function in $N_G$ points within the limits of the
 parameter space of the variable $t$. Since $N_G$ is very large, one can always
 approximate Eq. (2) with the following formulae:
 $$
  a_{ij}^2  = \left[ \sum_{m} S_i \ (u_0^k,t_E^l,t_m) \ F_m \right] \times  \Delta T
 $$
The above equation can be written $N_G$ times (for all the points in the
 grid). Since we have also $N_G$ unknown values $F_m$, we can write a full
 system of $N_G$ linear equations, which will allow to find the $N_G$ values
 $F_m$ which represents the function $F(t)$. 
 Thus the function F(t) exists for all values of $\rm { u_0} \ \rm and \ { t_E}
 $. \\\\
 Using the principal components
 decomposition of $\overline{ \tilde { S_i}}$ we can write:
$$ 
a_{ij}^2 \simeq \sum_{j=1}^{\rm N_C}  \left[ \int P_j \ (\rm t) \ F(\rm  t) \ dt \right] \ a_{ij} 
$$
Thus, finally we see that the distribution of the second moment is not more than a 
linear combination of the the $\rm N_C$ firsts moments (mean). Thus the mean
 of the second moment will not bring any additional information with respect to the
 mean. With a similar reasoning it is possible to show that the same property is
 true for the $N^{th}$ moment. \\\\
 Consequently we can conclude that all the information concerning the distribution
 of the $a_{ij}$ is contained in the mean of these components. No additional 
 information (uncorrelated information) will be found by looking at higher order 
 moments.
\subsection{From the principal components to the mass function.}
It is possible to re-express the previous definition of $\left<a_{j}\right>$
 by using the number density $\rho(\rm {{ u_0}},\rm {{ t_E}, \rm  M} )$ to observe a given amplification of parameters  $(\rm {{ u_0}},\rm {{ t_E}} )$ with a lens of mass M. In such case, the sum can be approximated very
 closely with an integral expression:
$$
  \left<a_{j}\right>  \simeq \int \int \int \rho(\rm {{ u_0}},\rm {{ t_E}, \rm  M} ) \ a_{j} \ (\rm { u_0},\rm { t_E})\ du_0 \ dt_E \ dM
$$
 If we define the efficiency function of the experiment, $\Theta \left(\rm { u_0},\rm {t_E}\right)$, the lensing rates for 1 solar mass lenses, $\Gamma_0(t)$
 and the mass function, $\phi(M)$ the formulae for $\bf{ \rho}$ reads:
$$
 \rho \left (\rm {{ u_0}},\rm {{ t_E}, \rm  M} \right) = \Gamma_0 \left(\frac{t_E}{\sqrt{M}}\right) \phi(M) \ \Theta \left(\rm { u_0},\rm {t_E}\right)
$$
 Leading to the following expression for $\left<a_{j}\right>$:
$$
 \left<a_{j}\right>  \simeq \int \int \int  \Gamma_0 \left(\frac{t_E}{\sqrt{M}}\right) \phi(M) \ \Theta \left(\rm { u_0},\rm {t_E}\right) \ a_{j} \ (\rm { u_0},\rm { t_E}) \ du_0 \ dt_E \ dM
$$
It is easy to re-arrange this integral in the following way:
$$
  \left<a_{j}\right>  \simeq \int \psi(M) \phi(M) \ dM
$$
With:
\begin{equation}
 \psi(M) = \int \int \Gamma_0 \left(\frac{t_E}{\sqrt{M}}\right) \Theta \left(\rm { u_0},\rm {t_E}\right) \ a_{ij} \ (\rm { u_0},\rm { t_E}) \ du_0 \ dt_E
\end{equation}
Thus we see that basically projecting the microlensing light curve on a basis
 of function and taking the statistical sum is equivalent to projecting the
  mass function itself on another function. Consequently, we see that a 
 projection in the space of the observation (the light curve) is directly 
 equivalent to a projection in the space of the mass function. The problem
 of finding the optimal set of projections for the observations, is solved by
 using the principal component method. Then all we have to do is to calculate
 the ``image'' of these principal components in the space of the mass function.
\section{MONTE-CARLO SIMULATIONS.}
\subsection{Introduction.}
 To illustrate this method and show how the mass function can be reconstructed,
 we will use a series of Monte-Carlo simulations. We will simulate microlensing
 events by selecting the parameters of the events according to the density
 distribution, $\rho(\rm {{ u_0}},\rm {{ t_E}, \rm  M} )$. One additional
 parameter we will have to select is the amplitude of the source. To get
 random variables which reproduces the distribution $\bf{ \rho}$ we need
 to decompose the problem. Basically $\rho(\rm {{ u_0}},\rm {{ t_E}, \rm  M} )$
 has 3 parts: \\\\
  {\bf - The rates}, $\Gamma_M(t_E)$:
 $$
 \Gamma_M(t_E) = \Gamma_0(\frac{t_E}{\sqrt{M}})
 $$ 
 For $\Gamma_0$ we will adopt the analytical expression given in
 Mao \& Paczy\`nski (1996). \\\\
 {\bf The efficiency function,}  $\Theta \left(\rm { u_0},\rm {t_E}\right)$: \\\\
For the efficiency we will adopt the following criteria: an event
 is detected if it has a minimum number ($N_m$) of data points above
 a 5 $\sigma$ threshold. For a given duration $t_E$ this criteria
 is simply equivalent to put a threshold on the impact parameter $u_0$.
 All we have to do is to search for the maximum value $u_T$ of $u_0$ such
 that the amplification of at least $N_m$ data points is above
 5 $\sigma$. If the sampling is even, with a time step $\delta t$,
 it just mean that the amplification
 must be larger than  5 $\sigma$ in a window of time around the maxima  $\Delta T=N_m \ \delta t$. Provided the maxima is at the origin of the time axis,
 it will finally result in the condition: 
\begin{equation}
 S_j\left(\rm {u_T},\rm {t_E},\frac{\Delta T}{2}\right) > 5 \ \sigma
\end{equation}
The distribution of $u_0$ itself is uniform, to get et set of $u_0$ values
 we will use a random generator which will provide a uniform distribution
 between 0 and the maximum value for $u_0$, $u_{M}(t_E)$. $u_{M}(t_E)$ 
 corresponds to the impact parameter for the brightest source given by Eq xx.
Once we have selected $t_E$ from the distribution $\Gamma_M(t_E)$, we will
 calculate $u_{0,thresh}$ using Eq xx, if our random $u_0$ value is above
 $u_{0,thresh}$, $\Theta \left(\rm { u_0},\rm {t_E}\right)=0$, otherwise
 $\Theta \left(\rm { u_0},\rm {t_E}\right)=1$. To summarize: \\\\
$$
{\Theta \left(\rm { u_0},\rm {t_E}\right)} = \cases{  
 1 & $u_0$ $<$ $u_T$ \cr
 & \cr
 0 & $u_0$ $>$ $u_T$ \cr}
$$
 {\bf The mass function, $\phi(M)$ :} \\\\
 In all our simulation we will adopt a pure power law expression for
 the mass function, with a lower ($M_{I}$) and an upper cut-off ($M_{S}$).
$$
{\phi(M)} = \cases{  
 M^{-\alpha} & $M_{I}$  $<$ M $<$ $M_{S}$ \cr
 & \cr
 0 & otherwise\cr}
$$
 {\bf The amplitude:} \\\\
 we will assume that the the flux of the un-amplified source amplitude \
 distribution is a power law with
 an exponent of -2 (Zhao, {\it et al.} 1995), in a given range of amplitude $A_{max}$ and
 $A_{min}$. In our simulation, the amplitudes will be generated by Monte-Carlo
 method, using this power law.
\subsection{Description of the implementation.}
To implement our Monte-Carlo simulation we will select the random variables
 in the following order: \\ \\
 {\noindent }  - The amplitude of the source, according to the probability law 
$A^{-\nu}$ \\
 {\noindent }  - The mass of the lens, according to the probability law 
$A^{-\alpha}$ \\
 {\noindent }  - The duration of the event $t_E$ using the distribution, $\Gamma_M(t_E)$ \\
 {\noindent }  - The impact parameter $u_0$ using an uniform distribution in the range, 0, $u_{M}(t_E)$.  \\
 {\noindent } - Once $t_e$ and $u_0$ are selected we apply the efficiency cut-off,
 using the function $\Theta \left(\rm { u_0},\rm {t_E}\right)$. \\\\
 This procedure produces a complete set of variables, $A$, $t_E$, $u_0$ for
 the events passing the efficiency cut-off. Form the variable we compute the
 microlensing light curve using Eq. 1. We add noise to the light curves according to the Poisson statistics. Using this procedure, it is possible to simulate
 a set of microlensing light curves $S_j$. 
\subsection{Example}
In this example we will take an amplitude range which is typical for
 microlensing images towards the Galactic Bulge: \\\\
 $A_{max}=10^6$, \ $A_{min}=10$ \\\\
 For the mass function we take the following parameters:
$$
{\phi(M)} = \cases{  
 M^{-2} &  0.05 $<$ M $<$ 10.0 \cr
 & \cr
 0 & otherwise\cr}
$$
Using these parameters we simulated several series of microlensing light 
 curves. For illustration, we extracted a 
 sample of light curves from one of these simulations, they are
  presented in Fig. 1. The parameters were
 adjusted in order that in a simulation the total number of microlensing events simulated be close to 100. Then we proceeded to the principal components decomposition. The orthonormal set of principal components was calculated using a
 singular value decomposition. The first 4 principal components are presented 
 in Fig. 2.

\begin{figure*} 
\centerline{\psfig{angle=0,figure=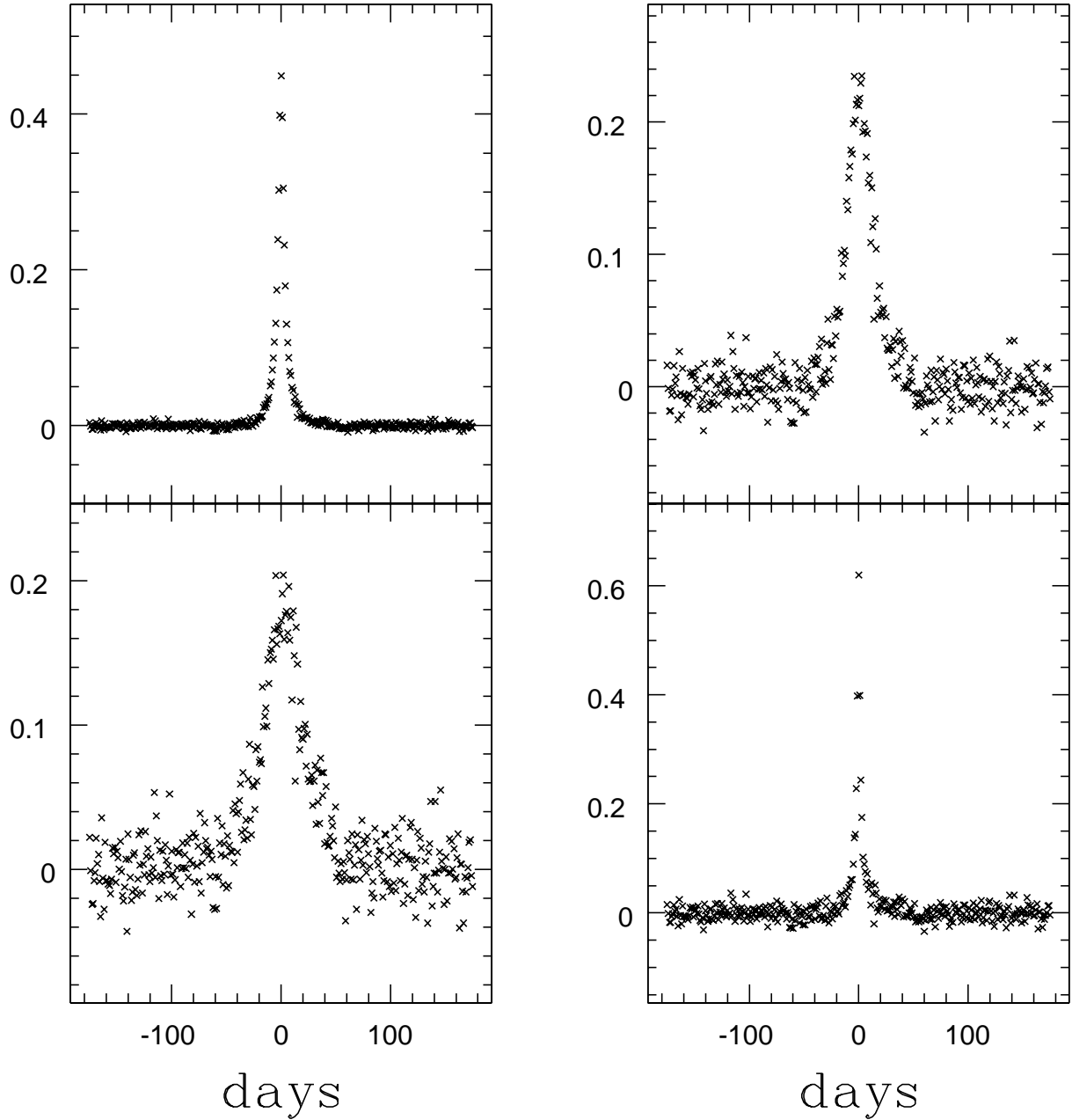,width=18cm}}
\caption{
 Example of light curves as obtained using the Monte-Carlo simulation procedure
 described above. 
  }
\end{figure*}
\begin{figure*} 
\centerline{\psfig{angle=0,figure=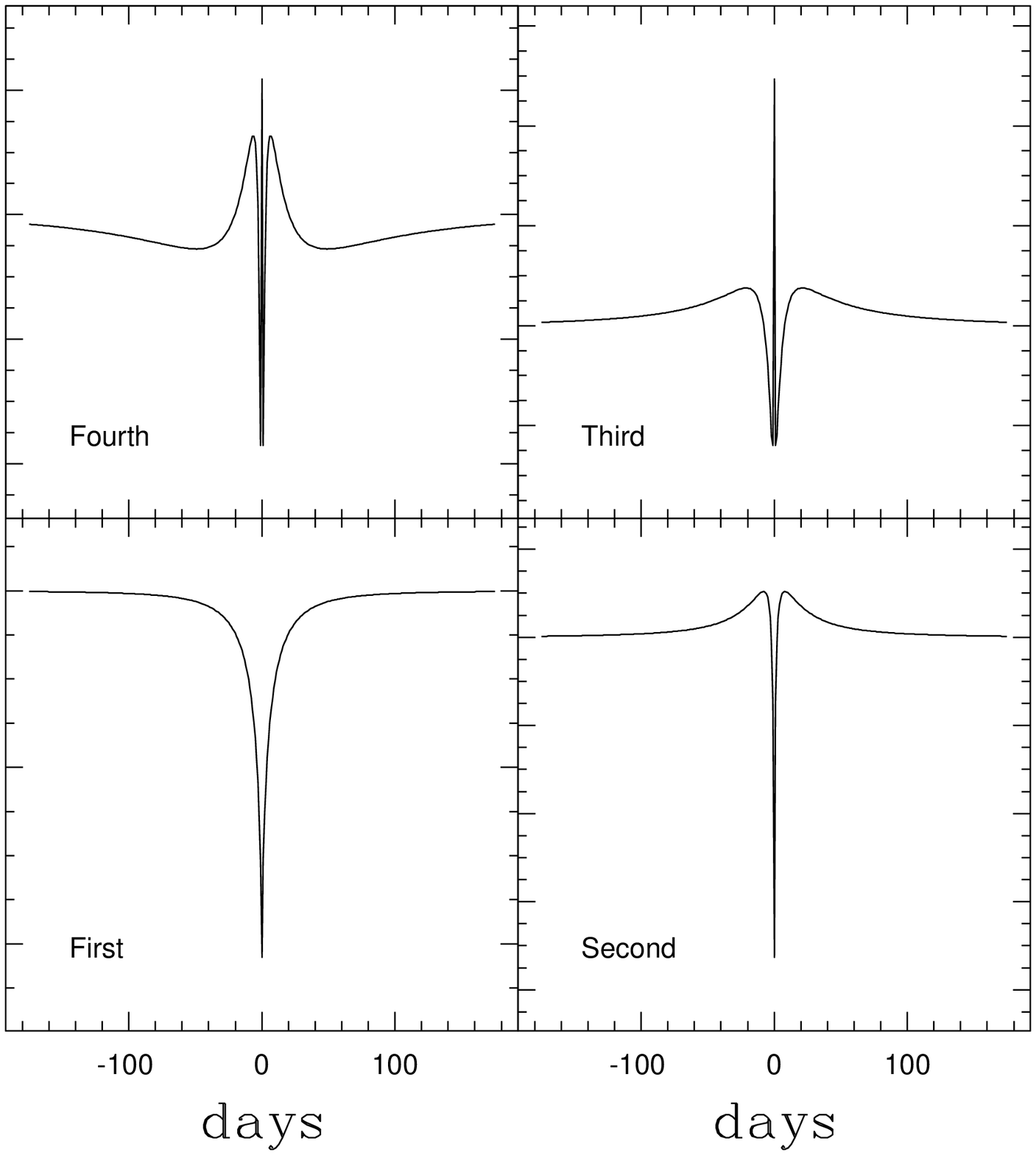,width=18cm}}
\caption{
The first 4 principal components corresponding to the simulation presented
 in Fig. 1. 
  }
\end{figure*}
The functions $\psi(M)$ corresponding to the principal components in the space of the mass
 function can be computed using Eq. 3. Projecting the data on the principal
 components is equivalent to projecting the mass function on $\psi(M)$. To illustrate
 our discussion, an example of $\psi(M)$ functions calculated using the settings of
  the previous Monte-Carlo simulation are presented in Fig. 3.
\begin{figure*} 
\centerline{\psfig{angle=0,figure=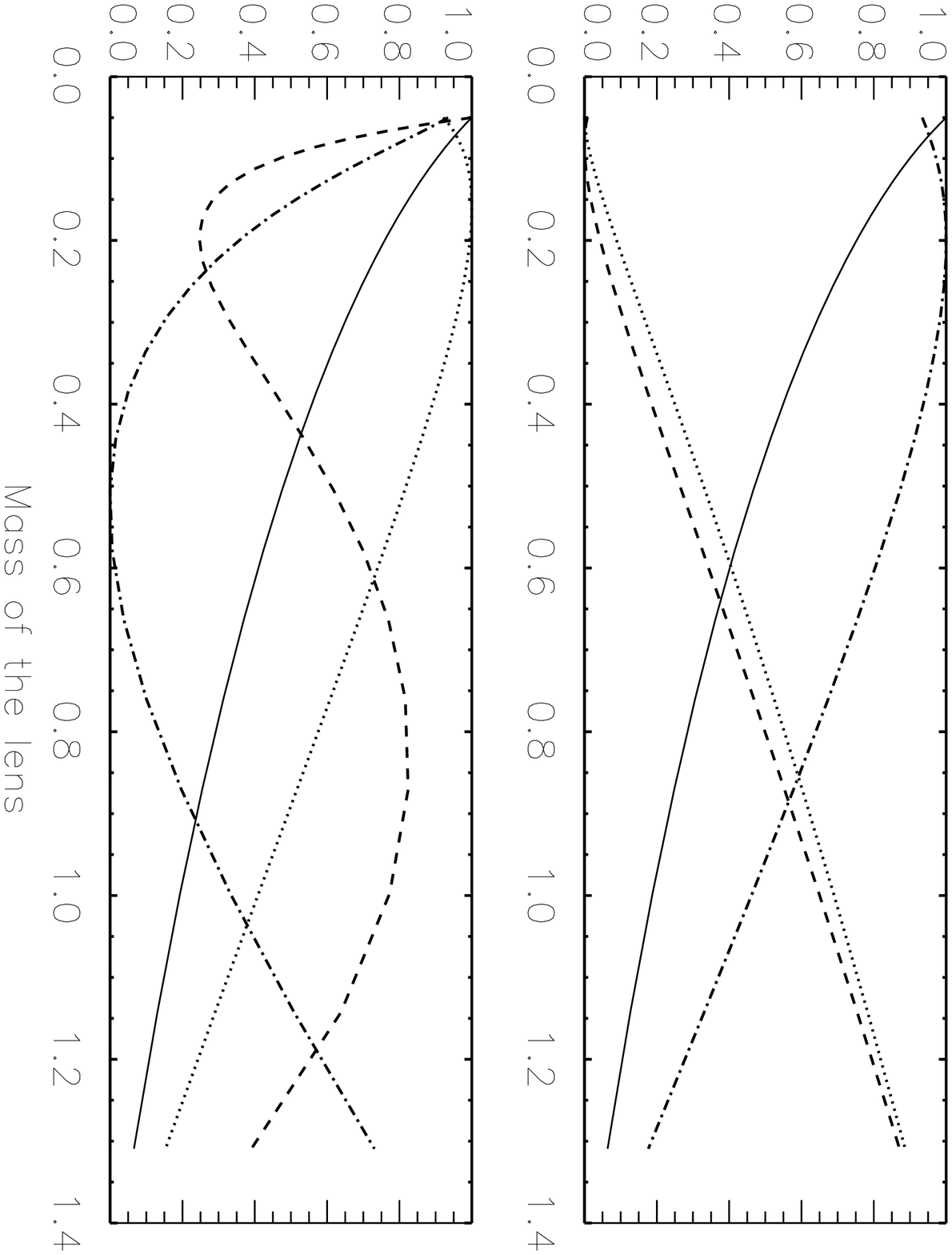,width=14cm}}
\caption{
The function $\psi(M)$ which corresponds to the principal components. In the
 right panel we present the 4 original function, while in the other
 panel we present an orthonormal linear combination of these 4 functions.
 (Note that in the right panel the function were scaled to have the same
 amplitude).
  }
\end{figure*}
\subsection{Estimating the mass function.}
To illustrate the ability of the method to reconstruct the mass function we 
 will assume that exponent of the mass the mass function,  $\alpha$ 
 is unknown. For each simulation we will calculate the 4 principal component
 of the set of microlensing light curves we have simulated. We will compute 
 the sum of the projection on a principal component of all the light curves 
 $<a_j>$.  Using Eq. 3
 we will calculate the function which corresponds to each principal components
 in the space of the mass function. Then all we have to do is to compare the 
 projection of a trial mass function with $<a_j>$. The trial mass function
 which match the 4 $<a_j>$ as close as possible (in a least-square tens) 
 will be the best mass function. This procedure was applied for 1000
  simulation, each time the program derived the best value of the exponent
 of $\alpha$. The histogram of the values of alpha for the 1000 simulations
 is presented in Fig. 4.
\begin{figure*} 
\centerline{\psfig{angle=0,figure=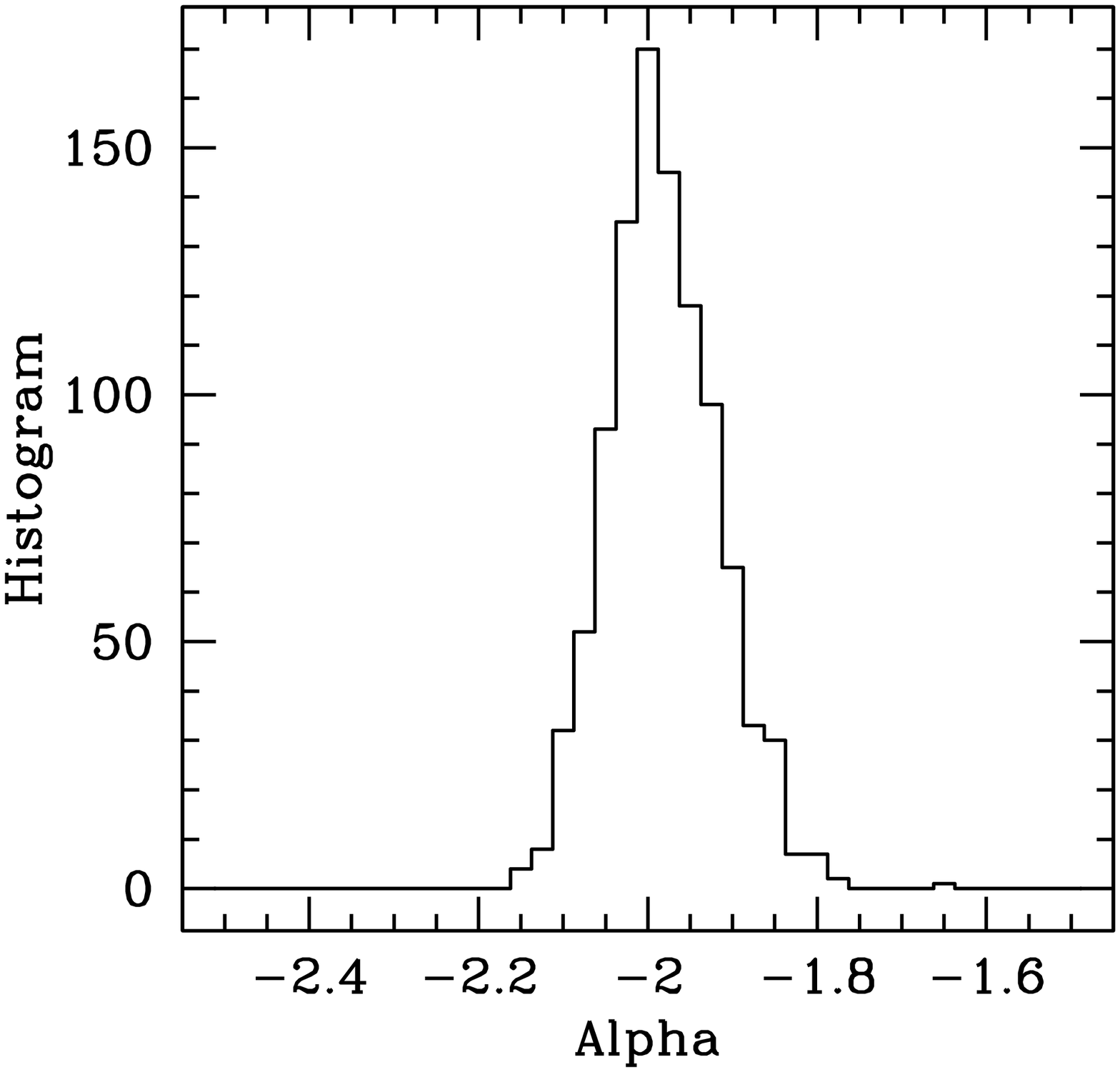,width=14cm}}
\caption{
 Histogram of the values of Alpha (the exponent of the mass function) found
 by fitting the Monte-Carlo simulation by projecting the data in the space 
 of the mass function.
  }
\end{figure*}
\label{lastpage}

\end{document}